\documentclass[12pt,oneside,a4paper,english]{article}
\usepackage[T1]{fontenc}
\usepackage[latin2]{inputenc}
\usepackage[margin=2.25cm,headheight=26pt,includeheadfoot]{geometry}
\usepackage[english]{babel}
\usepackage{listings}
\usepackage{color}
\usepackage{titlesec}
\usepackage{titling}
\usepackage{changepage}
\usepackage{amsmath}
\usepackage{hyperref}
\usepackage{enumitem}
\usepackage{graphicx}
\usepackage{fancyhdr}
\usepackage{lastpage}
\usepackage{caption}
\usepackage{tocloft}
\usepackage{setspace}
\usepackage{multirow}
\usepackage{titling}
\usepackage{float}
\usepackage{comment}
\usepackage{booktabs}
\usepackage{indentfirst}
\usepackage{lscape}
\usepackage{booktabs,caption}
\usepackage[flushleft]{threeparttable}
\usepackage[english]{nomencl}
\usepackage{xcolor}
\usepackage{siunitx}

\title{BIOREME Study Group Report} %
\makeatletter\let\Title\@title\makeatother

\newlist{steps}{enumerate}{1}
\setlist[steps, 1]{leftmargin=1.5cm,label = Step \arabic*:}

\begin{document}
\pagenumbering{roman} 

\pdfoutput=1
\begin{titlepage}
\begin{center}
\vspace{1.0cm}
\includegraphics[width=0.4\textwidth]{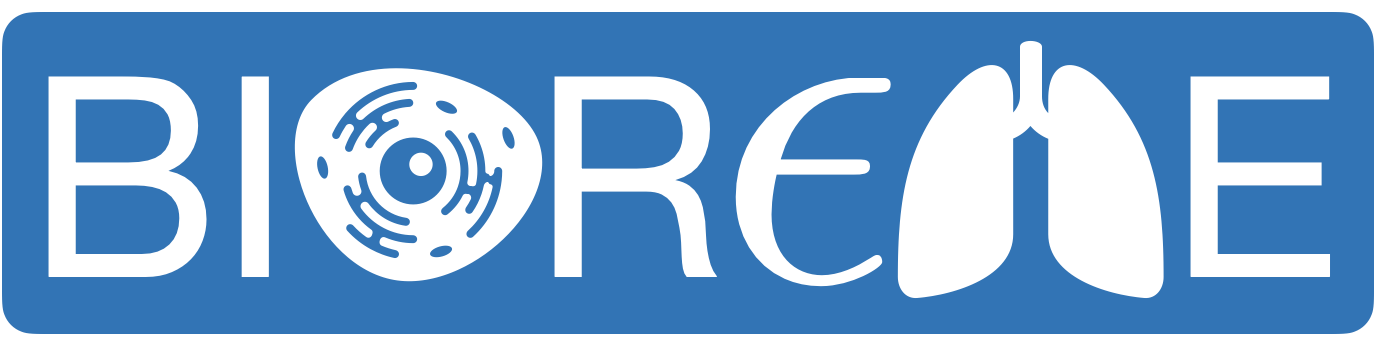}~\\[1cm]

{\large\textbf{A Mathematical Study Group in Data-Driven Biophysical Modelling for respiratory diseases with Industry and Clinicians}}
\vspace{.5cm}

\vspace{2.0cm}

\hrule
\vspace{.5cm}
{\huge\textbf{Using data collected from structured light plethysmography to differentiate breathing pattern disorder from normal breathing.}} %
\vspace{.5cm}

\hrule
\vspace{1.5cm}

\textsc{\textbf{Authors}}\\
\vspace{.5cm}
\centering

Bindi S. Brook -- University of Nottingham\\
Mathew Bulpett -- Oxford University Hospitals Trust\\
Robin Curnow -- University of Manchester\\
Emily Fraser -- Oxford University Hospitals Trust\\
Eric J. Hall -- University of Dundee\\
Shiting Huang -- University of Glasgow\\
Mariam Mubarak -- University of Glasgow\\
Carl A. Whitfield -- University of Manchester\\

\vspace{1cm}
*Corresponding author: 
General enquiries: \href{mailto:contact@bioreme.net}{contact@bioreme.net}

\vspace{2cm}

\centering Date of publication: \today %
\end{center}
\end{titlepage}

\newpage
\doublespacing
\renewcommand{\baselinestretch}{1}\normalsize
\tableofcontents
\renewcommand{\baselinestretch}{1}\normalsize
\thispagestyle{fancy} %

\newpage
\pagenumbering{arabic} 
\fancyfoot[C]{Page \thepage\ of \pageref{EndOfText}}

\section{Introduction} 
\label{sec:intro}

This report relates to a study group hosted by the EPSRC funded network, Integrating data-driven BIOphysical models into REspiratory MEdicine \href{https://www.bioreme.net/about}{(BIOREME)}, and supported by \href{https://www.softmech.org/}{SofTMech}  and \href{https://iuk.ktn-uk.org/industrial-maths/}{Innovate UK, Business Connect}. The BIOREME network hosts events, including this study group, to bring together multi-disciplinary researchers, clinicians, companies and charities to catalyse research in the applications of mathematical modelling for respiratory medicine. The goal of this study group was to provide an interface between companies, clinicians, and mathematicians to develop mathematical tools to the problems presented. The study group was held at The University of Glasgow on the 17 - 21 June 2024 and was attended by 14 participants from 8 different institutions. Below details the technical report of one of the challenges and the methods developed by the team of researchers who worked on this challenge.

\subsection{The Challenge}
\label{sec:challenge}

\textbf{Background:} Many people with long COVID suffer from breathlessness and breathing pattern disorders (BPD) \cite{EvansEtAl:2023sl}. Investigations of long COVID sufferers may be unremarkable, e.g., CT scans and lung function tests can appear normal. The EXPLAIN Study (ongoing \cite{grist_hyperpolarized_2021}; this project is linked to challenge holders) has identified possible sources of lung damage in long COVID patients that cannot be detected with routine tests. Currently, BPD diagnosis typically follows after the exclusion of other conditions and disorders. Understanding the mechanisms that drive breathlessness and BPD is crucial for accurate and timely diagnosis.

``Normal'' breathing typically follows a ``rectangular'' pattern of inhale time to exhale time in a ratio of $1:2$. This ratio is approximate and will differ from person to person (the quoted figure is based on a theoretical/average optimal use of lung capacity). As breathing pattern focuses on the ratio of inhale to exhale time, normal breathing potentially includes breathing that would be classified as shallow or deep. We note that some patients with diagnosed BPD will have completed a breath training program and may adopt a rectangular breathing pattern when observed. Other subjects who do not have breathing pattern disorder exhibit ``box breathing'', which is common in yoga training where the practice is to inhale for a $4$ count, hold a breath for $2$, and exhale for $6$, leading to a pattern of inhale to exhale of $1:1$. Thus, it is a challenge to distinguish irregular breathing patterns from inhale and exhale data alone. 

In assessing BPD, the challenge holders indicated clinicians typically focus on imbalances between chest (thoracic) and abdominal breathing. This is done by physically examining patients while breathing; the physician or therapist places a hand on the chest and a hand on the abdomen to assess any imbalance.  

Structured light plethysmography (SLP) is a relatively new technology (ca. 2010) that enables contactless (i.e. non-invasive) measurements of breathing activity. Specifically, SLP measures estimated volume changes (\emph{plethysmography}), using image processing techniques \cite{deBoer:2010slp}. A grid of light (i.e., \emph{structured light}) is projected onto a subject's chest and abdomen while in a seated or supine position. An array of cameras track grid pattern intersection points over time. Calibration algorithms are then used to reconstruct the line-of-sight displacement of the thoracoabdominal wall, which can be used to calculate proxy measures for lung volume over time. SLP has shown promise for measuring chest wall motion and asynchrony between regions of the thoracoabdominal wall, for example, for post-operative patients \cite{ElshafieEtAl:2016sl} and COPD patients \cite{MotamediFakhrEtAl:2017sl}. Other, more invasive, tests to measure breathing flow rates and lung volume are possible.

\textbf{The problem:} Given SLP, demographic, and clinical measurements for a cohort of study participants and controls, 
\begin{enumerate}
    \item determine variables that can be used to differentiate normal breathing from BPD, and to better understand the range of ``normality'', and
    \item explore the data to better understand lung mechanics and diaphragmatic movement.
\end{enumerate}
That is, can SLP be used together with demographic and clinical data to detect inefficient breathing? If so, which variables are most useful? Are any variables confounders? Secondarily, can the data be used to understand the sensitivity of the measurements to chest/abdomen division placement (the current division is fixed at the xiphisternum)?

\textbf{Data available:} The challenge holders made available demographic, clinical and SLP measurements from healthy controls ($n = 31$) 
and patients with BPD ($n = 67$) 
both at rest and after exercise (1-minute sit-to-stand test). The SLP data was collected using a PneumaCare Ltd SLP device, and summary statistics were generated using the PneumaView 3D software. 

Demographic variables included gender, age, and date of first clinic appointment. Non-SLP clinical variables included a Dyspnoea-12 Score and a Nijmeng Score (hyperventilation score). We note that there were significant gaps in the demographic and clinical provided. In particular, not all subjects had gender recorded, and the clinic did not record Nijmeng scores for all patients as clinicians found it to be insensitive measure of breathing pattern disorder. Longitudinal measurements were only available for some patients; these were removed for subsequent analysis so that there was only one measurement per patient (the first measurement). 

Summary SLP measurements include the thoracic to abdominal movement ratio during tidal breathing and following exercise, regularity of breathing pattern, inspiratory to respiratory ratio, and respiratory rate. Raw SLP time series displacement data (chest, abdominal and total) was also made available after this was identified by the study group as potentially interesting to analyze.

\begin{table}[h!]
\begin{tabular}{@{}lcl@{}}
\toprule
Variable                & Unit         & Description                                                 \\ \midrule
Resp Rate               & \unit{\minute^{-1}}   & respiration rate (breaths per minute)                       \\
$T_i$                   & \unit{\second}         & duration of inhalation                                      \\
$T_e$                   & \unit{\second}         & duration of exhalation                                      \\
$T_{\rm tot}$           & \unit{\second}         & duration of whole breath (inhalation + exhalation)          \\
$d_{\rm chest}$         & \unit{\percent}        & chest displacement, as percentage of total displacement     \\
$d_{\rm abdom}$         & \unit{\percent}         & abdomen displacement, as percentage of total displacement   \\
IE$_{50}$               &          & ratio of inspratory to expiratory flow at 50\% tidal volume \\
Breath Phase Angle      & \unit{\degree}     & phase difference between chest and abdomen displacements    \\ \bottomrule
\end{tabular}
\caption{SLP summary statistics produced automatically by the PneumaView 3D software.}
\label{tab:indices}
\end{table}

\subsection{Proposed Solution}
\label{sec:solution}

Our proposal was to approach the problem in 3 stages:
\begin{enumerate}
    \item Apply statistical analysis methods initially to understand the features of the summary data (or lack thereof). 
    \item Characterise the temporal and spatial information in the data to see if any features were being masked by looking solely at summary information. 
    \item Consider physiological models to account for individual differences and attempt to fit mechanistic parameters of BPD.
\end{enumerate}

\section{Methods} 
\label{sec:methods}

\subsection{Dimensionality reduction and cluster analysis}
\label{sec:dim_reduction}

\textbf{Linear methods.} Principal Component Analysis (PCA) is an embedding approach that creates linear combinations of quantitative variables, which can be ordered by the amount of the total variance each explains within a data set (see, e.g., the review \cite{JolliffeEtAl:2016rv}). This is achieved by finding the eigenvectors and eigenvalues of the covariance matrix of a dataset. The set of eigenvectors (principal components) forms a new basis for the data, with each row of the eigenvectors representing the contribution of each input parameter to it. The corresponding eigenvalues indicate the amount of variance explained by each principal component. A similar embedding can be achieved for qualitative variables using a multiple correspondence analysis (see, e.g, the review \cite{Greenacre:2010rv}). Such an analysis considers the covariance of the complete disjunctive table, a matrix representation of ones and zeros indicating whether the levels of each category (as columns) are present or not for each observation. Factor analysis of mixed data (see \cite{Pages:2014fa}) is an approach that roughly combines PCA for qualitative variables with multiple correspondence analysis for qualitative variables. The resulting embedding can be used for visualization and data processing and can be undertaken using the FactoMineR package in R \cite{LeEtAl:2008rp}.

\textbf{Nonlinear methods.} Dimensional reduction of the summary statistics was used to embed them onto 2-dimensional space and to analyse qualitatively any clustering between participants. This was done using a Uniform Manifold Approximation and Projection (UMAP) \cite{mcinnes_umap_2018}. UMAP is a nonlinear dimensionality reduction algorithm that first forms probability distributions for each sample in high dimensional parameter space based on the distance between that sample and its $n$ nearest neighbours (NN), where $n$ is user-defined. Samples are then iteratively moved in the embedding space, starting from random locations, until similar probability distributions to those in high dimensional space are found.  The input values were z-score normalised before performing the UMAP analysis using the umap package in python \cite{mcinnes_umap_2018}.

Both the PCA and UMAP were performed on the following 5 summary statistics \{Resp Rate, $d_{chest}/d_{abdom}$, Breath Phase Angle, $IE_{50}$ $T_i/T_e$\}. $T_{tot}$ was not included in this analysis as it was expected to be highly correlated with Resp Rate. FAMD was performed on all of the available tidal (pre-sit-stand test) breathing summary statistics \{Resp Rate, $d_{chest}$, Breath Phase Angle, $IE_{50}$, $T_{tot}$, $T_i$, $T_e$, $T_i/T_e$, $T_i/T_{tot}$, Gender, Dyspnoea-12\}. 

\subsection{Time series analysis}
\label{sec:time_series}

It has been noted in previous studies using SLP measurements \cite{alhuthail_measurement_2021} and other pulmonary function tests \cite{frey_temporal_2011} that measures of temporal variability may be as or more informative about disease state than average measurements. For example, when looking for a breathing pattern disorder, it may be that the average breath profiles of each subject are idiosyncratic, but that variations around this individual baseline indicate something about the underlying breathing pattern. Therefore, we extracted breath-by-breath data from the time-series data by first identifying breaths in and out in the time series, then computing $T_i$, $T_e$, $T_{\rm tot}$, $d_{\rm chest}$, and $d_{\rm abdom}$ for each breath. 

Breath detection and identification is performed automatically within the PneumaView 3D software, however the cannot be exported. So, we developed an algorithm to achieve separate breaths in the time series displacement data (over the whole grid). This algorithm worked as follows.
\begin{enumerate}
    \item The displacement data was smoothed using a Savitzky-Golay filter with a window of 10 points and polynomial order 2.
    \item Each of the smoothed data points was given ``exhaling'' or ``inhaling'' status based on the sign of the gradient (first-order central difference) at that point.
    \item The start and end times of all inhalations and exhalations were recorded at all the points with a status different to the last data point. This was then used to calculate the duration and ``volume'' (i.e. cumulative displacement) of each inhalation and exhalation.
    \item Any inhalations or exhalations with a volume less that 25\% of the median over the whole test were absorbed into the previous breath (i.e. they were treated as small deviations that were actually still part of the previous breath).   
\end{enumerate}

From this breath-by-breath data we then computed the following two measures of variability for each index
\begin{align}
{\rm QCV}_f &= \frac{Q_3(f) - Q_1(f)}{Q_2(f)},\\
{\rm BBV}_f &= Q_2({\rm BBV}_{f,i}) \; {\rm where} \; {\rm BBV}_{f,i} = \frac{|f_{i+1} - f_{i}|}{\bar{f}},
\end{align}
where $f \in \{ T_i, T_e, T_{\rm tot}, d_{\rm chest}, d_{\rm abdom}\}$ is the index in question, $i \in {1, \ldots, N}$ is the breath number, and $\bar{f} = \sum_i f_i / N$ is the mean of $f$ over the whole test. The functions $Q_1(x)$, $Q_2(x)$, and $Q_3(x)$ indicate the first, second and third quartiles of $x$ respectively. Thus, the measure QCV$_f$ is simply the quartile-based coefficient of variation of $f$ over the whole test. BBV$_f$ is a measure of breath-to-breath variability, i.e. the average relative varaition of an index from one breath to the next.

\subsection{Spatial analysis}
\label{sec:methods_spatial}

By default, the PneumaView 3D software labeled pixels as chest or abdomen using a horizontal dividing line across the center of the measured area (xiphisternum placement). Custom definitions of chest and abdomen could be made within the software by selecting two subsections of pixels and the data for these selections could then be exported for further analysis. Since custom selections could only be made and exported for one participant at a time, only a single alternate definition for chest and abdomen was investigated and only for 8 participants with severe breathlessness.

\begin{figure}[h!]
    \centering
    \includegraphics[width=0.5\textwidth]{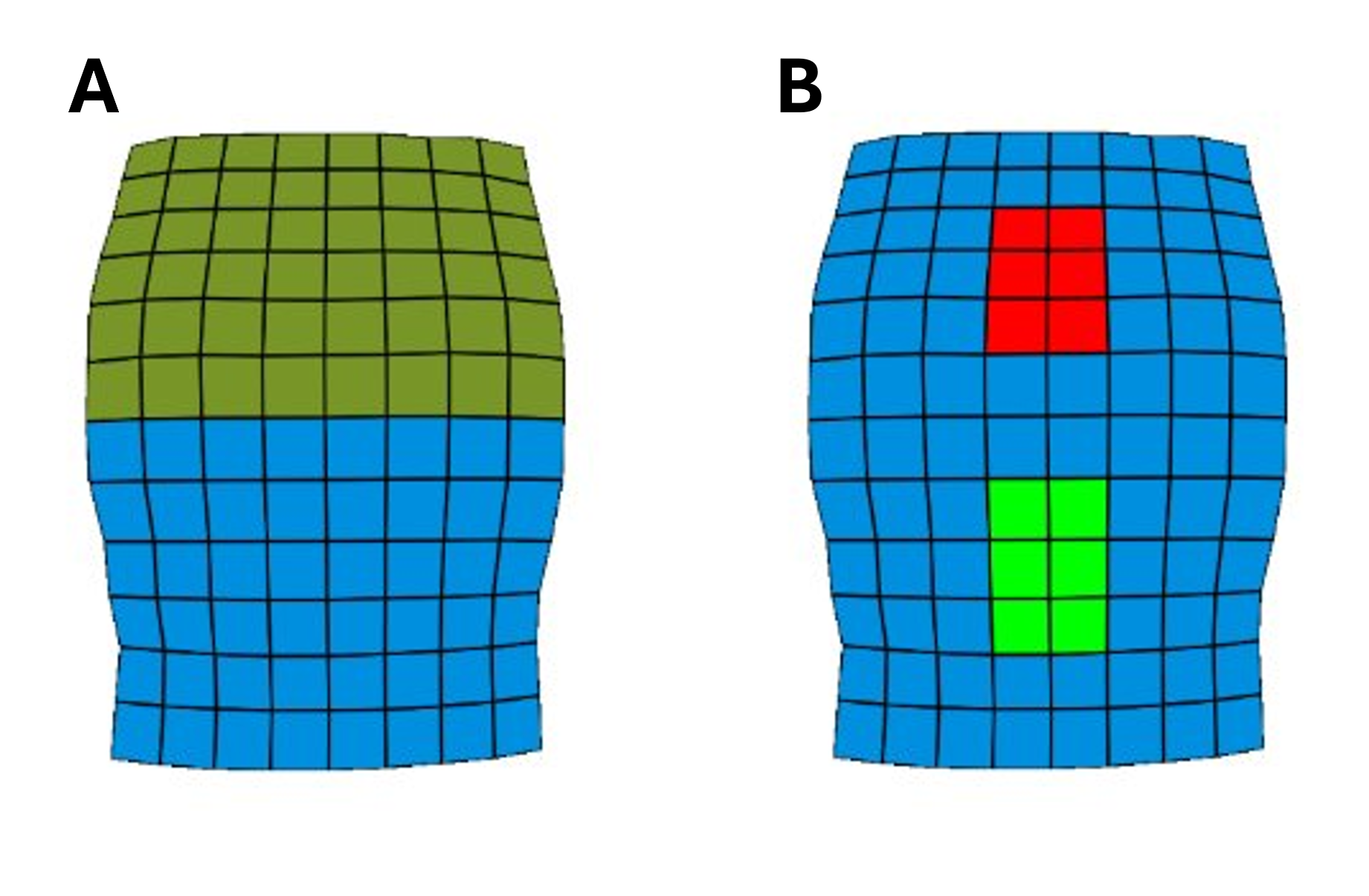}
    \caption{\textbf{A}) default definition of the chest/abdomen split through the center of the measured area. \textbf{B}) custom selection of chest and abdomen.}
    \label{fig:chest_abd_split}
\end{figure}

The custom selection was made to attempt to reduce the potential confounding effect of lateral movement on the edges of the measured area. For each of the 8 participants, areas 2 pixels wide and 3 tall were selected for the chest and abdomen, leaving one unselected pixel above and below the horizontal center line of the measured area (Figure~\ref{fig:chest_abd_split}). 

Furthermore, for one control group patient, displacement data for individual pixels were also extracted from the PneumaView 3D software. This single sample of pixel-by-pixel displacement data was used to test the feasibility of extracting other spatial features of the breath pattern. One clear limitation of this data, compared to optoelectronic plethysmography \cite{aliverti_compartmental_2001}, is that the displacements are measured at fixed points in the plane, and not material points. Furthermore, unlike the SLP prototype detailed in \cite{deBoer:2010slp}, the field of view does not extend across the whole torso, which precludes using registration methods to create 3D deformation maps and infer the motion of material points. 

Nonetheless, if we consider the surface $z = f(x,y)$ that is created by the relative displacement $z$ of each pixel in the grid, then the evolution of the shape of this surface over time may reveal some aspects of breathing pattern lost in the averaging process which is related to BPD. In this case, we calculated the Laplacian operator applied to $f$
\begin{align}
\label{laplace} L = f_{xx} + f_{yy},
\end{align}
and the Gaussian curvature of the surface $f$
\begin{align}
\label{gcurv} G = \frac{f_{xx}f_{yy} - f_{xy}^2}{(1 + f_x^2 + f_y^2)^2}.
\end{align}

To compute the derivatives, we first compute $z$ by splitting the data into breaths, as outlined in the previous section. Then for each pixel, the inhaled and exhaled displacements and calculated as
\begin{align}
    d_{\rm inh}^{(n)} &= D(t_{\rm inh,end}^{(n)}) - D(t_{\rm inh,start}^{(n)})\\
    d_{\rm exh}^{(n)} &= -D(t_{\rm exh,end}^{(n)}) + D(t_{\rm exh,start}^{(n)})
\end{align}
where $D(t)$ are the raw displacement values for each pixel, and $t_{\rm inh,start}^{(n)}$, $t_{\rm inh,end}^{(n)}$, $t_{\rm exh,start}^{(n)}$, and $t_{\rm exh,end}^{(n)}$ are the start and end times for the $n$th inhalation and exhalation respectively. Then we define
\begin{align}
    z = \mathrm{median}(d_{\rm inh}^{(n)} + d_{\rm exh}^{(n)}).
\end{align}
We found that, in this case, it made little difference to the spatial pattern of the final results whether we consider the mean instead of the median, or whether we used the inhaled or exhaled displacements instead the sum of the two. To compute the gradients, we used central differencing, except at boundaries where we used backwards and forwards differencing for upper and lower boundaries respectively. For, second order gradients, the differencing operators were applied twice. Unit grid-spacing in both directions was assumed.

\section{Results} 
\label{sec:results}

\subsection{Exploratory Data Analysis}
\label{sec:data_analysis}

\begin{figure}[h!]  %
    \centering
    \includegraphics[width=\textwidth]{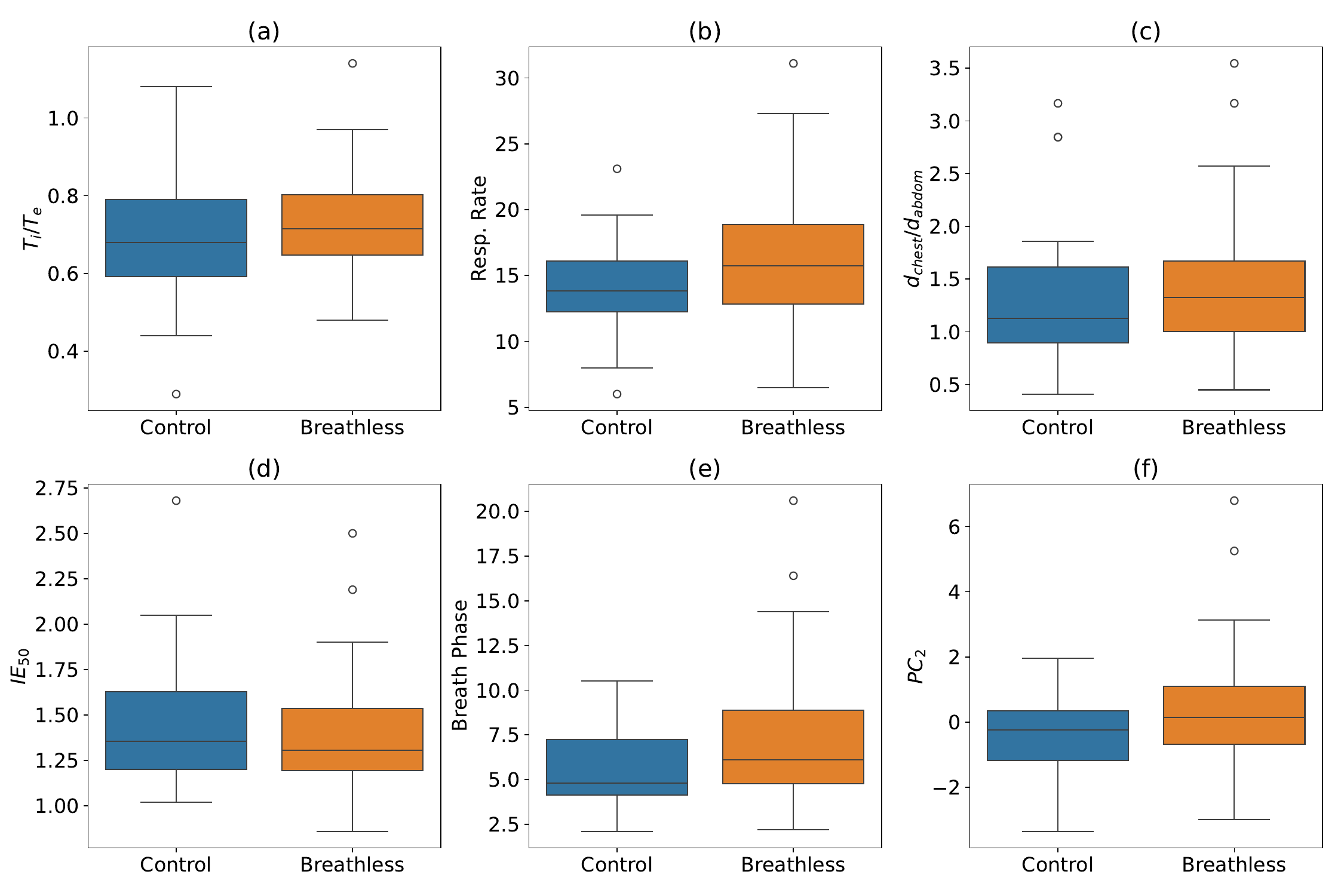}
    \caption{Box plots of 5 summary statistics from the SLP measurements for the breathless and control groups: (a) $T_i/T_e$ the ratio of mean inspiration duration to mean expiration duration; (b) $RR$ the respiratory rate; (c) $d_{\rm chest}/d_{\rm abdom}$ the ratio of mean chest displacement to mean abdomen displacement; (d) $IE_{50}$ the ratio of inspiratory to expiratory flow at 50\% tidal volume; (e) Breath Phase Angle: the phase difference between the chest and abdomen displacements. (f) shows the variable PC2, which is a linear combination of breath phase angle and respiratory rate.}
    \label{fig:control_v_breathless}
\end{figure}

\begin{figure}[h!]
    \centering
    \includegraphics[width=\textwidth]{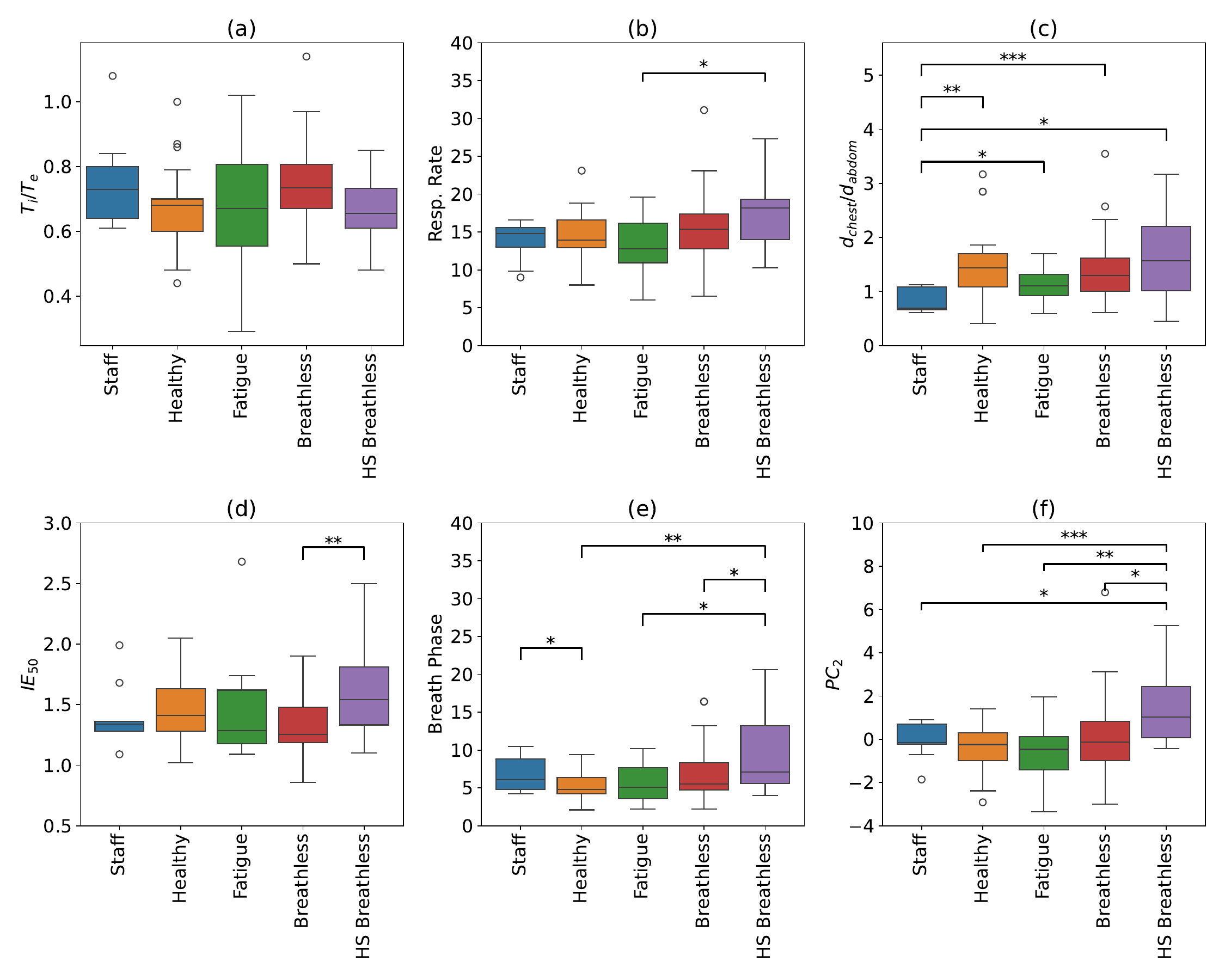}
    \caption{Box plots of 5 summary statistics from the SLP measurements for the breathless and control subgroups, as labelled: (a) $T_i/T_e$ the ratio of mean inspiration duration to mean expiration duration; (b) $RR$ the respiratory rate; (c) $d_{\rm chest}/d_{\rm abdom}$ the ratio of mean chest displacement to mean abdomen displacement; (d) $IE_{50}$ the ratio of inspiratory to expiratory flow at 50\% tidal volume; (e) Breath Phase Angle: the phase difference between the chest and abdomen displacements. (f) shows the variable PC2, which is a linear combination of breath phase angle and respiratory rate.}
    \label{fig:subgroups}
\end{figure}

The quantitative summary statistics listed in Table~\ref{tab:indices} showed no clear differences between breathless and control groups (Figure~\ref{fig:control_v_breathless}). Separating into smaller subgroups there we some notable differences but nothing that consistently indicated breathing pattern disorder, see Figure~\ref{fig:subgroups}. Breath phase angle was significantly different in the "high-severity breathlessness" group compared to the other controls (except the staff group). However, this group only consisted of $n = 11$ subjects, and given that many t-tests were computed across all subgroups and summary statistics, these significant differences could be due to chance variation. Therefore, this would need to be tested in a larger controlled study. 

\begin{figure}[h!]  %
    \centering
    \includegraphics[width=\textwidth]{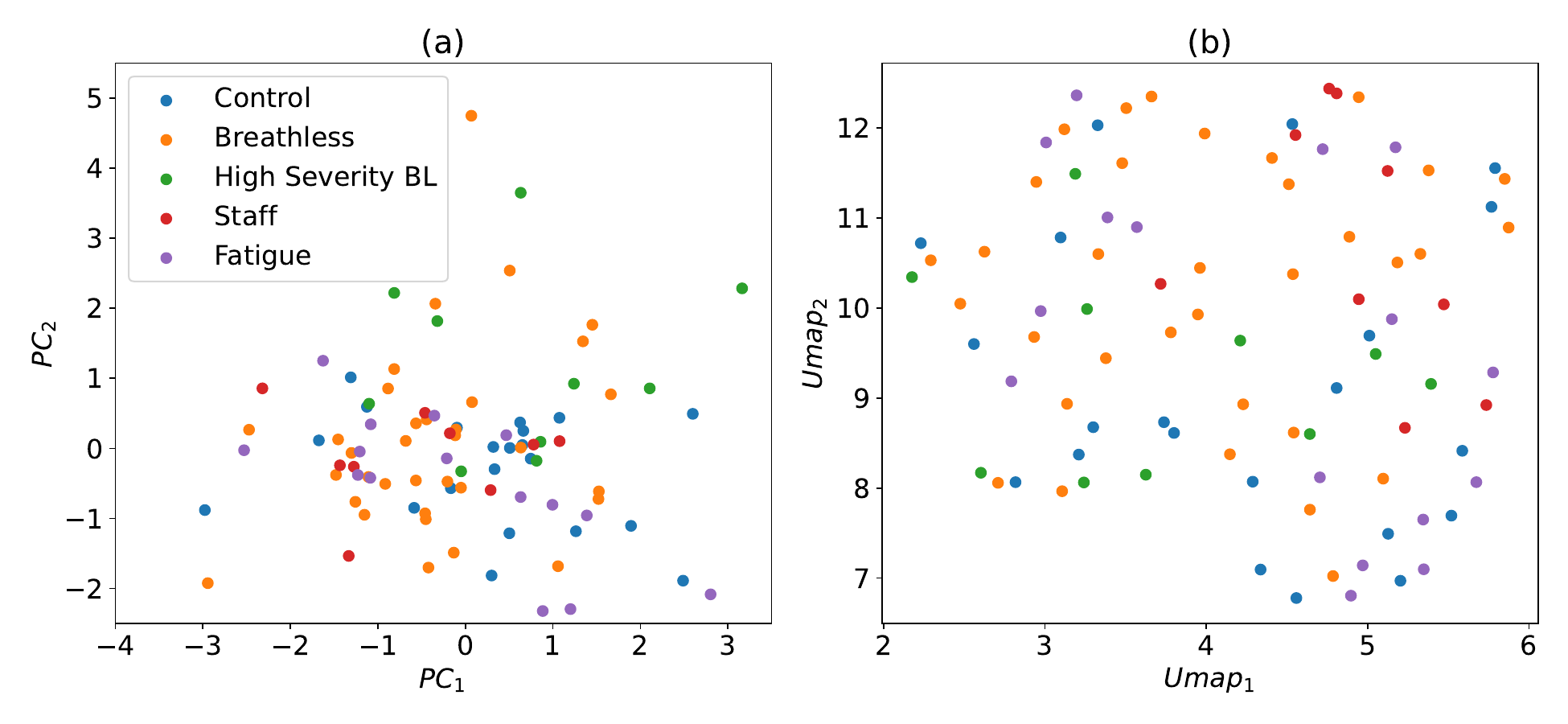}
    \hfill{}
    \includegraphics[width=0.9\textwidth]{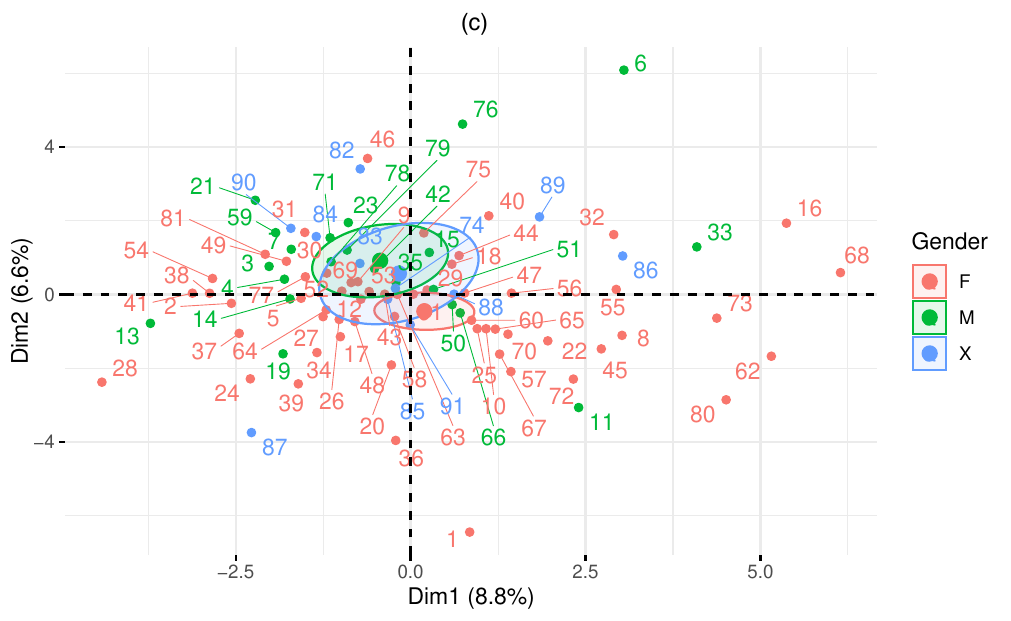}
    \caption{(a) PCA and (b) UMAP of Resp Rate, $d_{\rm chest}/d_{\rm abdom}$, Breath Phase Angle, IE$_{50}$ and ratio of $T_i$ and $T_e$ for each of the EXPLAIN study participants with severe breathlessness (n = 8), non-severe breathlessness (n = 28) and fatigue but no breathlessness (n = 15) as well as the healthy controls from the study (n = 20) and the staff members who also provided SLP measurements (n = 10). (c) FAMD, considering quantitative variables above and additional qualitative variables: Gender and Dyspnoea-12 for the EXPLAIN study participants (n = 50) and control (n = 21).}
    \label{fig:PCA-UMAP-FAMD}
\end{figure}

Embedding of summary statistics across breaths for each study participant using PCA, UMAP, and FAMD revealed no clear differences between participants (Figures~\ref{fig:PCA-UMAP-FAMD}). The PCA and UMAP used the following 5 summary statistics \{Resp Rate, $d_{chest}/d_{abdom}$, Breath Phase Angle, $IE_{50}$ $T_i/T_e$\}. %
PC1 and 2 account for approximately 62\% of the variation (with PC3 taking this to 82\%), therefore this does not indicate a substantial reduction in dimensionality from the original 5. Scatter plots of the first two components of the PCA and the UMAP projections are shown in figure \ref{fig:PCA-UMAP-FAMD}.

It is worth noting that the second component of the PCA did exhibit some separation of the high-severity breathless group from the rest of participants. This eigenvector of this component consisted almost entirely of the breath phase angle and respiratory rate in equal measure, so we constructed a new index consisting of a linear combination of the two measures by shifting each by the mean and normalising by the standard deviation in this dataset,
\begin{align}
PC2 = \frac{RR - 14.8}{4.3} + \frac{BP - 6.5}{3.3}.
\end{align}
Here $RR$ is the respiratory rate and $BP$ is the breath phase angle. Figure~\ref{fig:subgroups}(f) shows that this does separate the high-severity group from the rest, however the same statistical caveats apply so this would need to be verified in a larger dataset.

\begin{figure}[h!]
    \centering
    \includegraphics[width=0.49\textwidth]{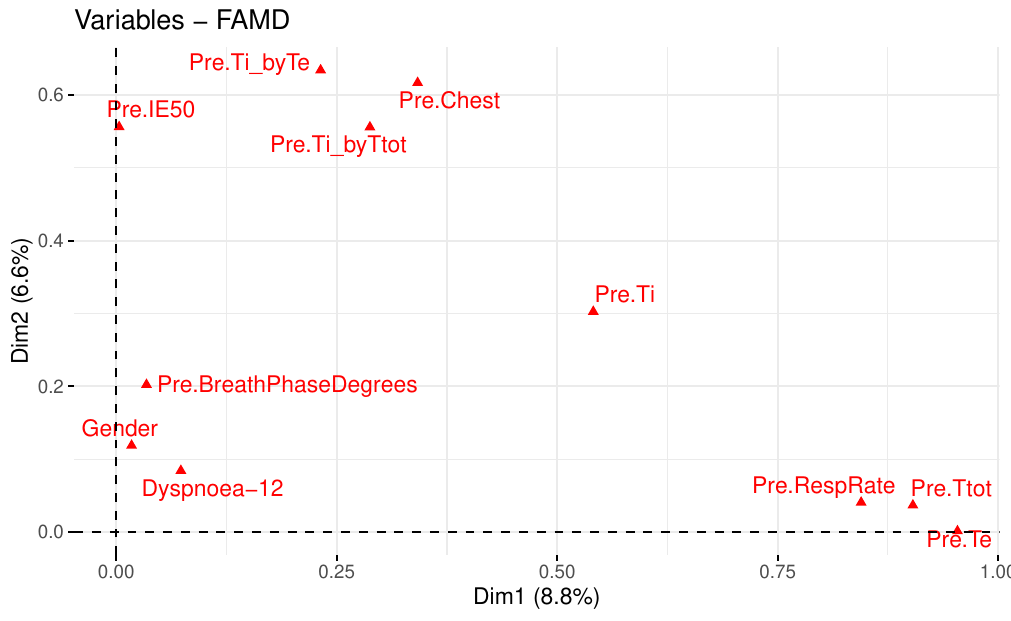}
    \hfill{}
    \includegraphics[width=0.49\textwidth]{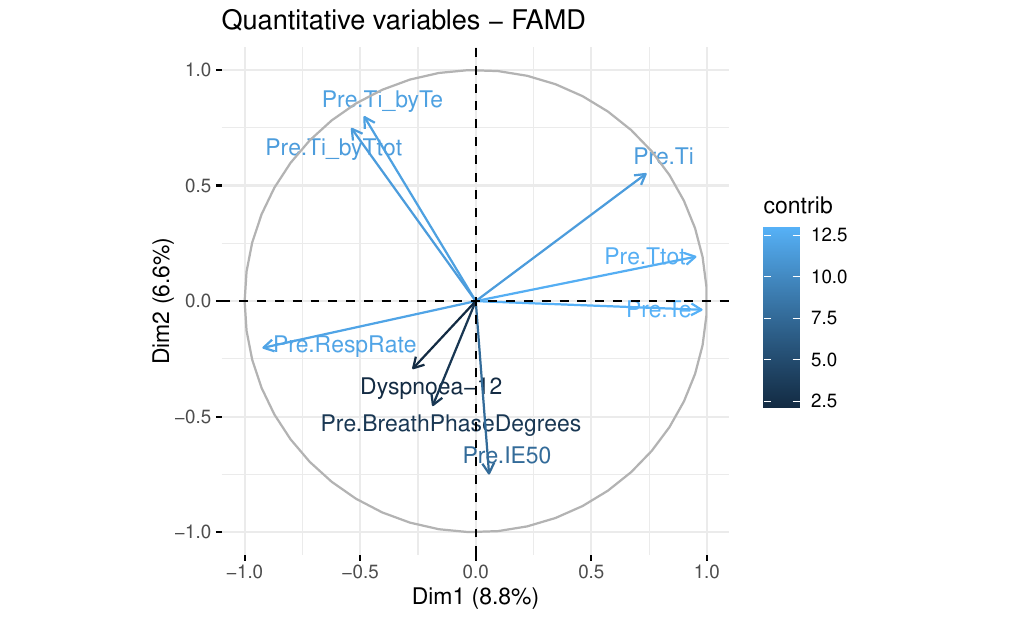}
    \caption{(Left) Graph of variables considered in the FAMD analysis, highlighting correlation between variables and the principal dimensions. (Right) Correlation circle for quantitative variables, with a heatmap indicating contributions to each dimension. The first two principal dimensions only account for 8.8\% and 6.6\% percent of the explained variation in the data.}
    \label{fig:FAMD}
\end{figure}

The FAMD analysis includes the qualitative and quantitative variables in \ref{fig:FAMD}. The first two principal FAMD dimensions account for only 8.8\% and 6.6\% of the total variation, respectively, indicating that the FAMD analysis does not lead to a substantial dimensionality reduction. Subsequently, projection on the first two principal dimensions, \ref{fig:PCA-UMAP-FAMD}, yields no apparent clustering. 

\subsection{Breath-to-Breath Variability}
\label{sec:b2bvar}

\begin{figure}[ht]
\centering
\includegraphics[width=\textwidth]{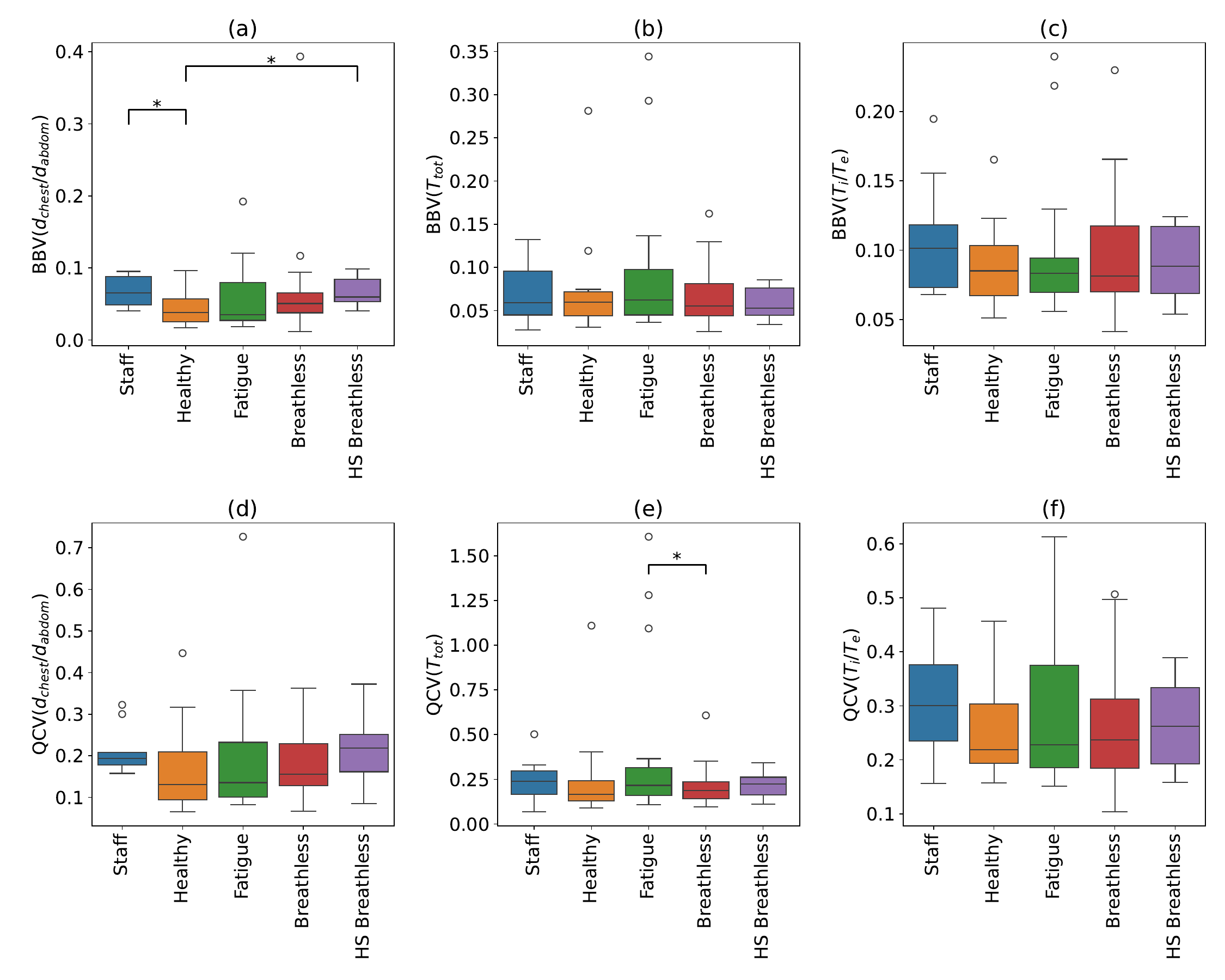}
\caption{Box plots of the breath-to-breath variability (BBV) of (a) the chest-to-abdomen displacement ratio, (b) breath duration, (c) inspiration to expiration duration ratio, split by participant subgroup. (d) (e) and (f) show the quartile coefficient of variation (QCV) for the same metrics respectively.}
\label{fig:BBV_QCV}
\end{figure}

Initially, we chose to look at the breath-to-breath variability in the ratio of $d_{\rm chest} / d_{\rm abdom}$, due to its perceived importance for diagnosis of BPD. Interestingly the high-severity breathless group did show significantly higher BBV in this parameter than the control group, but there was not a significant difference with other groups. However, exploring BBV for other parameters revealed no clear differences so it would appear that breath-to-breath variability does not identify BPD any better than the summary statistics in this dataset (figure \ref{fig:BBV_QCV}).

Overall heterogeneity measured by QCV also failed to reveal any clear differences, for any variable. This suggests that temporal variability is not a good marker of BPD, at least for SLP data. 

\subsection{Spatial analysis}
\label{sec:spatial}

\begin{figure}[ht]
    \centering
    \includegraphics[width=0.8\textwidth]{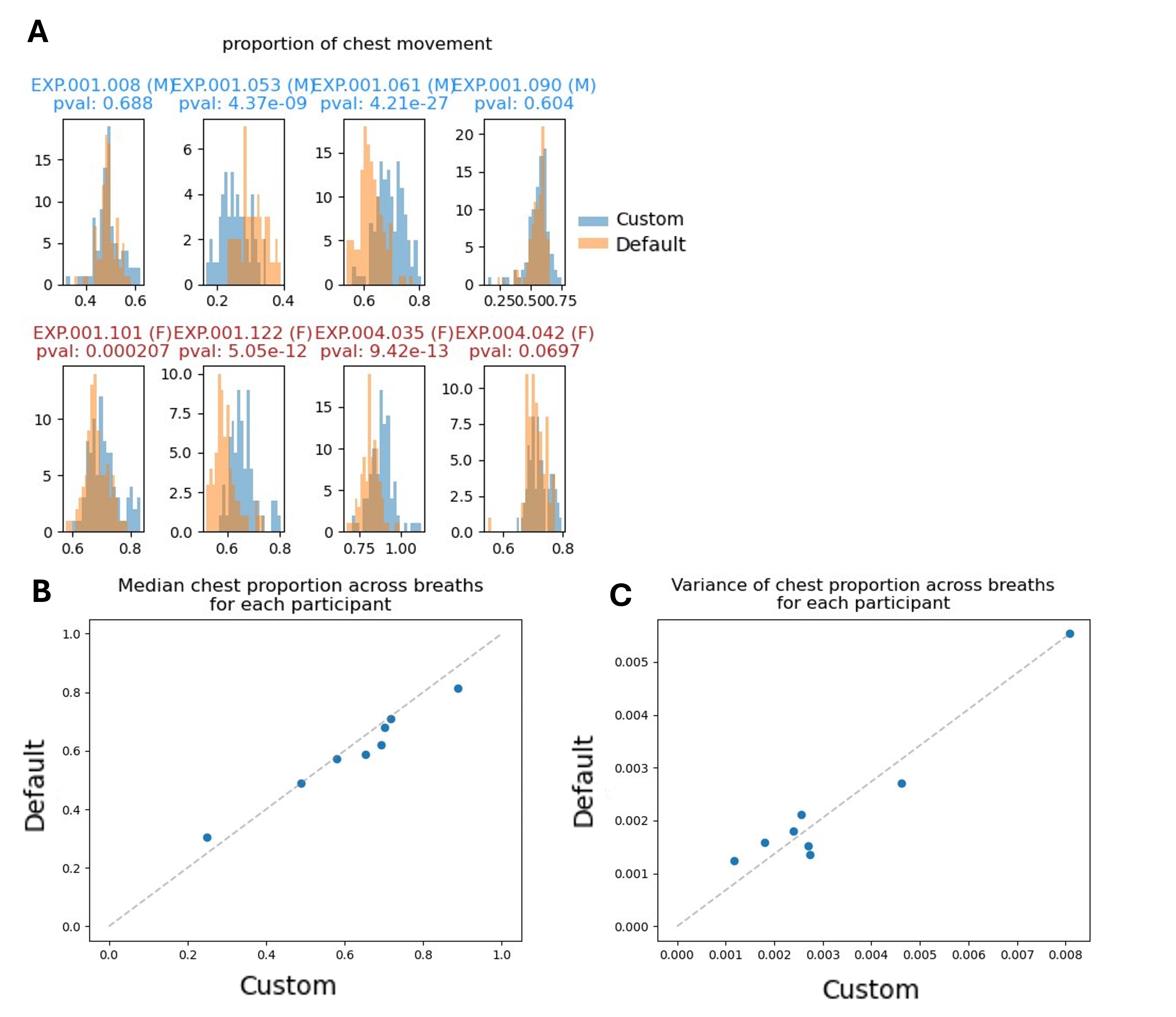}
    \caption{\textbf{A}) Histogram showing the distribution of $d_{chest}$ for each participant. For each participant, $d_{chest}$ is measured using a custom selection of pixels to define the chest and abdomen and with the default selection which splits the measured area in half (section \ref{sec:methods_spatial}). Median (\textbf{B}) and variance (\textbf{C}) across all breaths in each participant for the custom definition of chest and abdomen against the same measures for the default definition. Grey dotted line indicates equal median/variance for both definitions.}
    \label{fig:chest_abd_defn}
\end{figure}

Comparisons were made between the two definitions for chest and abdomen displacement for participants from the EXPLAIN study with severe breathlessness. For each participant, t-tests across all breaths were performed for each participant, with 5/8 having p-values $<0.001$ (Figure~\ref{fig:chest_abd_defn} \textbf{A}). Meanwhile, the median and variance of $d_{chest}$ across the breaths for the custom selection did not appear to shift far from the values given by the default selection (Figure~\ref{fig:chest_abd_defn} \textbf{B} and \textbf{C}).

\begin{figure}
    \centering
    \includegraphics[width=\linewidth]{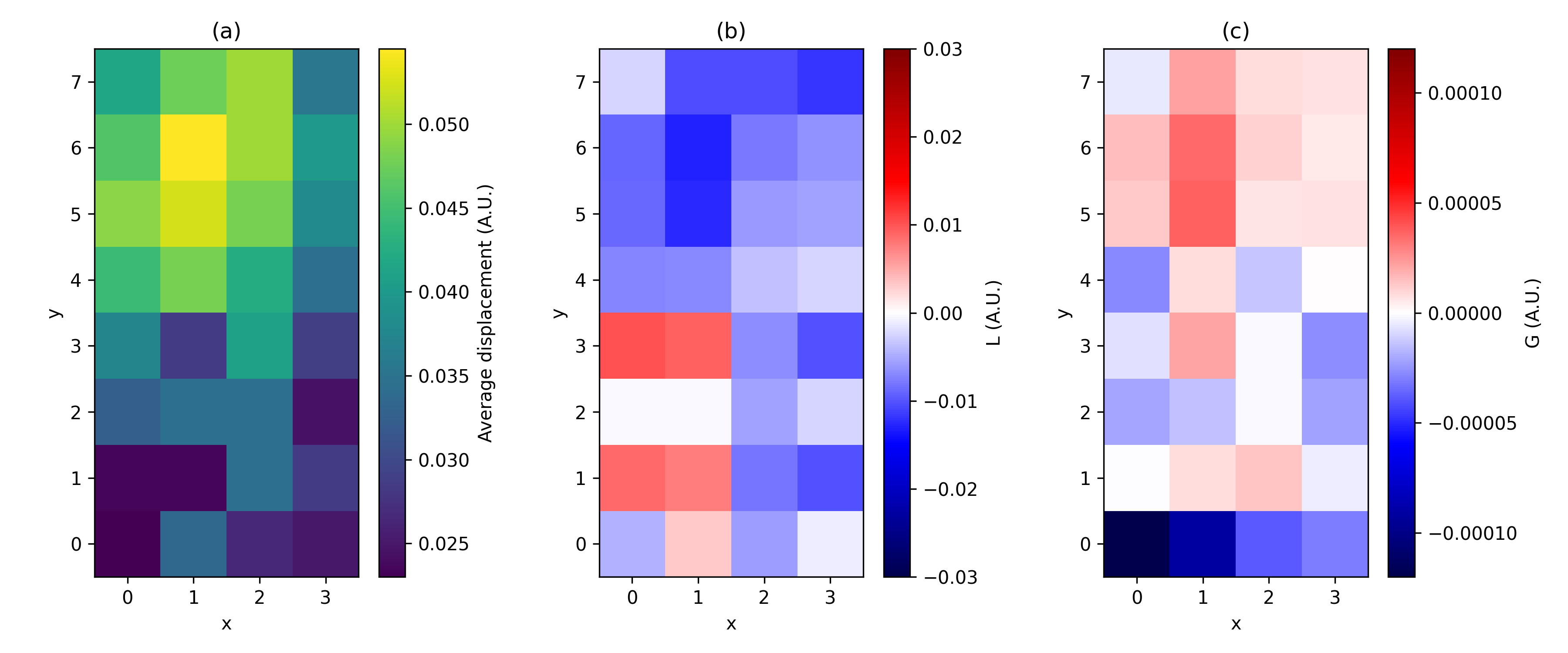}
    \caption{(a) Median pixel displacement over the course of a breath ($d_{\rm inh} + d_{\rm exh}$. (b) Result of applying the Laplacian operator, as described in equation \eqref{laplace} to the pixel displacements. (c) Result of applying the Gaussian curvature operator, as described in equation \eqref{gcurv} to the pixel displacements.}
    \label{fig:pixel_by_pixel}
\end{figure}

The results of processing the individual sample of a complete set of pixel displacements is shown in figure \ref{fig:pixel_by_pixel}. Little can be concluded from a single case, but in this example there is a clear difference between the displacements in the upper and lower halves, and fairly strong coherence of displacements within those halves. The results of applying the Laplacian operator in \ref{fig:pixel_by_pixel}(b) highlight regions where the displacements are concave ($L > 0$) and convex ($L < 0$). The chest displacements are found to be convex, while there is a mixed picture in the abdominal region. The Gaussian curvature reveals a similar picture but has a slightly more subtle interpretation. Positive values (seen in the chest region) represent more sphere-like (i.e. convex in all directions) deformations, while negative values (seen in the abdominal region) show hyperbolic-like deformations (convex in one direction but concave in the other). Zero values can either represent flat, linear or cylindrical deformations (i.e. where the surface is plane-like in at least one direction). We highlight these measures as generic ways of analysing the shape of the deformation surface, although it is impossible from one datapoint to say whether they will be informative. We posit that greater lateral deformation (e.g. ``shoulder breathing'') could alter the patterns of deformation seen in these maps, although this would have to be verified against measurements of deformation of the whole torso as in \cite{aliverti_compartmental_2001, deBoer:2010slp}.

\section{Discussion} 
\label{sec:disc}
Exploratory data analysis methods revealed no clear differentiation between participant groups with or without BPD. The modest proportions of the variation explained by each PCA/FAMD component, despite the removal of correlated variables from the data, possibly suggest a low signal-to-noise ratio within the data. Univariate investigations showed some differences between groups and especially between the severely breathless group and the others. This was particularly noticeable when using a linear combination of RR and BP angle. It is important to note, however, that repeated statistical testing and low sample sizes may overstate the true population differences between groups for these variables.

Two measures of temporal variability were used to try and differentiate participants with BPD. Although BBV provided a statistically significant difference between severely breathless participants and healthy controls, this difference was not seen between any other pair of groups. This analysis is also subject to the same issues of repeated statistical testing as with the summary statistics, suggesting it may be no better at identifying individuals with BPD, at least when measured in the manner seen in this data set. 

Defining an alternate definition for the chest and abdomen altered the distribution of some participant's data, although the median and standard deviations across all breaths were largely unchanged. It may be that other alternate definitions or investigating differences for all of the available participants would highlight differences more clearly. However, the software constraints made this impractical to do in the time available for the study group. Newer software versions will allow for custom definitions for multiple participants to be exported simultaneously, making investigating this more practical. It is also possible that with a wider field-of-view, the SLP measurements would be able to measure full thoracic deformation, enabling better estimates of chest and abdomen expansion as well as enabling other spatial analyses, such as estimating strain rates and displacements in the measurement plane through 3D shape reconstruction and registration methods.

In the summary statistics, it appears that there were no distinctions between breathless and control groups, which could indicate that there is variation within each group. This variability could be due to natural differences in breathing patterns among individuals, including the healthy controls. For example, Fleck et al. (2019) observed different ``regular'' breathing patterns in their control participants depending on amplitude and frequency of TA displacement \cite{fleck2019}. Similarly, Marconi and De Lazzari (2020) describe ``normal'' breathing patterns that can vary among individuals depending on pleural pressure waveform  \cite{Marconi_De_Lazzari_2020}. These variations among healthy participants could be substantial enough to overlap with those of BPD patients, posing a challenge in identifying BPD patterns.

Gender is typically considered to play a role in respiratory disorders. For example, Jin et al. (2020) found that compared with women, men have higher COVID-19 mortality rates \cite{jin2020}, suggesting that the mechanisms affecting respiratory patterns and outcomes may differ by gender. However, our data analyses did not identify any gender-based differences concerning breathing pattern disorders. Such a difference might be seen with an improved study design or a larger sample size in general. 

\section{Next steps}

Below, we suggest two possible mathematical routes that might be considered in future work. Firstly, more advanced approaches to analysing temporal and spatial data might be considered to determine variables that can be used to differentiate normal breathing from BPD and to better understand the range of breathing ``normality''. Secondly, a mechanistic model of the lungs that captures chest and abdominal breathing imbalances might provide insights into lung mechanics and diaphragmatic movement that might inform studies of BPD. The chances of success of each possible route might be improved with additional data and by ensuring the data provided is cleaned and complete. 

\textbf{Further temporal and spatial analyses.} Although the initial exploratory data analysis failed to directly identify summary variables that might be used to distinguish breath disorders from controls, it pointed to two types of functional data that might be amenable for further analysis. In particular, measures of temporal variability in breath-to-breath data should be further explored. As a first step, the properties of such signals, such as the presence of persistence or anti-persistence in the non-periodic variations, might be analysed (using, e.g., Hurst rescaled-range analysis \cite{MandelbrotWallis:1969lm} or a fractionally integrated model \cite{GrangerJoyeux:1980lm,Hosking:1981fd}). Although such exploratory analyses might be applied to presently collected data, a suggestion for future data collection might be to concentrate on capturing longer breath-to-breath sequences (as opposed to capturing breathing over a fixed time) to create observational records that are more comparable. 

The dimensionality reduction and clustering analysis considered in this report is applied to summary variables. Another logical next step might be considering similar exploratory analyses on more complex data objects, including time series and shape data. Object oriented data analysis frameworks (see, e.g., \cite{MarronDryden:2021oo}), might be useful for understanding modes of variation appropriate for differentiating SLP time series and related shape data. For example, one might imagine a cluster analysis based on the functional behaviour of breath-to-breath time series or the imbalance in chest-to-abdominal breathing.

\textbf{Mechanistic model for breathing imbalances} 
To diagnose BPD effectively, it might be essential to model both thoracoabdominal coordination, airway-lung mechanics, and the muscle activity driving the breathing process. Since breathlessness in BPD might attributed to how the diaphragm and intercostal muscles are used,  mechanistic compartmental models might best employed to roles of these muscle in breathing. Models such as Ward's two-compartment model focus on the rib cage distortions and the mechanical interactions between the diaphragm and intercostal muscles\cite{Ward1992}. However, it lacks the ability to simulate pressure changes within the lungs and airways and therefore may not provide a mechanistic explanation for the breathlessness patients experience. 

On the other hand, Marconi and De Lazzari's airway-lung mechanics model simulates the respiratory system's dynamic behaviour by considering airflow, airway resistance, and pleural pressure changes \cite{Marconi_De_Lazzari_2020}. This model include detailed equations for the pleural pressure, chest wall pressure and respiratory muscle pressure, but does not account for the rib cage and abdominal motion. 

By combining these two models we might be able to simulate the thoracoabdominal imbalances (from Ward's model) and evaluate their effects on airflow and pressure dynamics (from Marconi's model). Specifically, as the rib cage and diaphragm move during breathing, pressure variations within the lungs and airways can be modelled more accurately, accounting for mechanical distortions and physiological responses, and perhaps muscle activity
Furthermore, we can integrate real-time SLP data on chest and abdominal movements into the model. This would allow us to quantify these mechanical distortions and link them to physiological consequences such as inefficient airflow or abnormal lung pressure.

According to \cite{Marconi_De_Lazzari_2020}, during inspiration, the respiratory muscles generate a pressure, $P_{mus}$, which counteracts the pressure within the chest wall, $P_{cw}$, resulting in negative pressure within the pleural cavity, $P_{pl}$. Given that the external pressure, $P_{ref}$, is considered to be zero, this negative pleural pressure allows air to flow into the respiratory system. Conversely, during expiration, $P_{cw}$ exceeds $P_{mus}$, causing $P_{pl}$ to become positive, thereby expelling air from the respiratory system.\\
\begin{figure}[h!]
    \centering
    \includegraphics[width=0.7\linewidth]{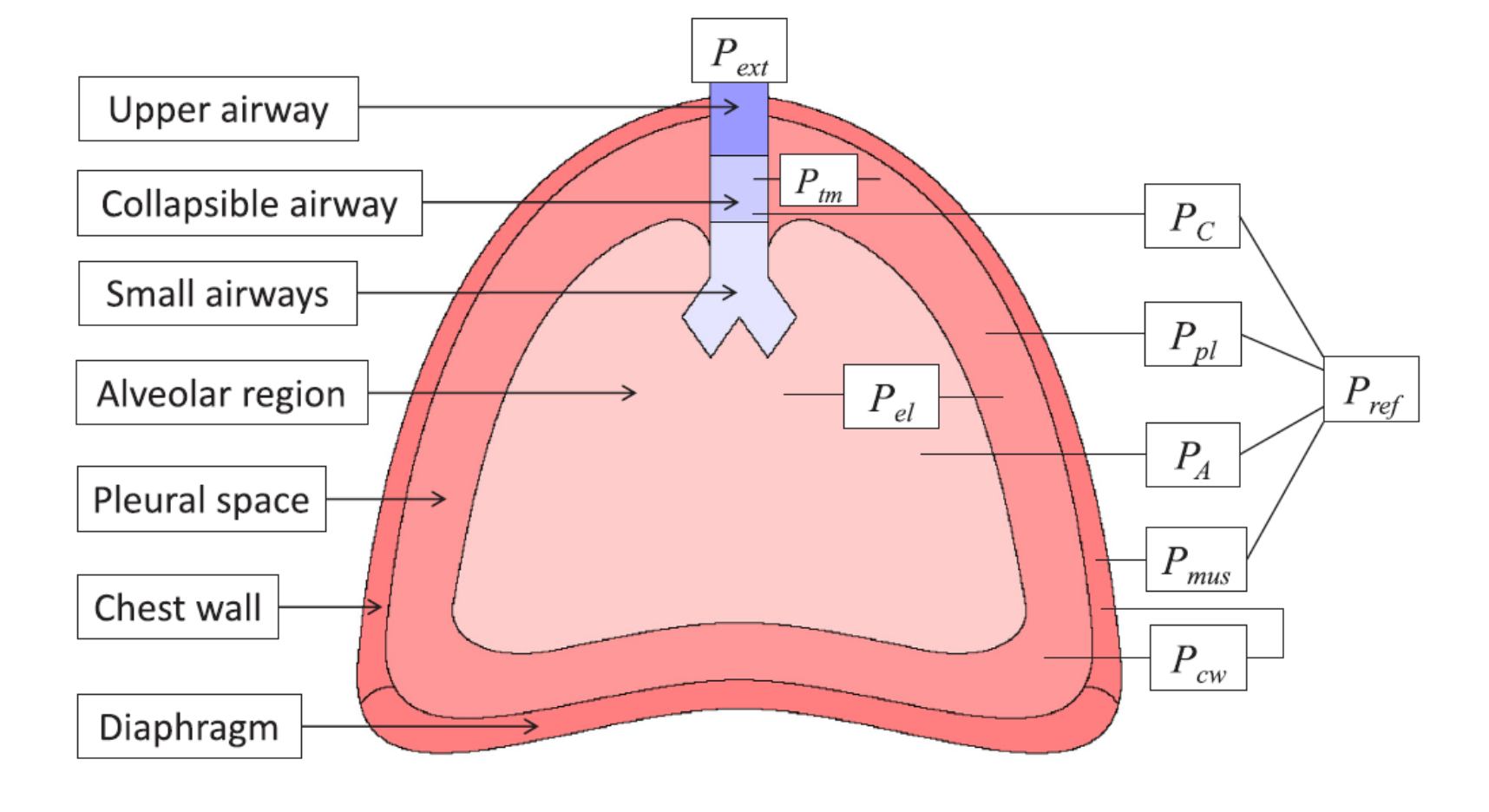}
    \caption{The schematic of the respiratory anatomical structure and pressure. The $P_{ext}$ and $P_{ref}$ represent the reference pressure, it is always seen as zero. $P_{pl}$ represents the pleural pressure. $P_{tm}$ is the transmural pressure, it represents the pressure drop between collapsible and pleural space. $P_{cw}$ and $P_{mus}$ are the pressure of chest wall and respiratory, respectively. $P_{A}$ and $P_{C}$ represent the pressure in alveolar and collapsible airways \cite{Marconi_De_Lazzari_2020}.}
    \label{respiratory}
\end{figure}

To differentiate between abdominal and chest breathing, it is essential to distinguish the physiological roles of the respiratory muscles. The original muscle pressure, \( P_{\text{mus}} \), can be specifically separated into the pressure generated by the diaphragm, \( P_{\text{dia}} \), and the pressure generated by the intercostal muscles (muscles between the ribs), \( P_{\text{inter}} \). The total pressure source is the sum of \( P_{\text{dia}} \) and \( P_{\text{inter}} \). Different breathing types (chest or abdominal) can be simulated by specifying different pressure variations in \( P_{\text{dia}} \) and \( P_{\text{inter}} \).\\

The combined model, enhanced with SLP data, might allow us to captures both the mechanical distortions and the pressure dynamics of breathing. This could offer a personalised diagnostic tool that analyses muscle activity and temporal changes in BPD.

\section*{Author Contributions}
\addcontentsline{toc}{section}{Author Contributions}

Bindi S.~Brook, Robin Cunrow, Eric J.~Hall, Shiting Huang, Mariam Mubarak, and Carl A. Whitfield contributed equally to the preceding analysis and authorship of this report. The challenge holders, Mathew Bulpett and Emily Fraser, contributed numerous insights and data, and their contributions are greatly appreciated. 

\label{EndOfText}
\newpage
\pagenumbering{Roman} 
\fancyfoot[C]{Page \thepage\ of \pageref{EndOfDoc}}

\addcontentsline{toc}{section}{List of Figures}
\listoffigures

\newpage
\addcontentsline{toc}{section}{References}
\bibliography{document.bib} 

\begin{thebibliography}{10}

\bibitem{EvansEtAl:2023sl}
R.~Evans, A.~Pick, R.~Lardner, V.~Masey, N.~Smith, and T.~Greenhalgh, ``Breathing difficulties after covid-19: a guide for primary care,'' {\em BMJ}, vol.~381, 2023.

\bibitem{grist_hyperpolarized_2021}
J.~T. Grist, M.~Chen, G.~J. Collier, B.~Raman, G.~Abueid, A.~McIntyre, V.~Matthews, E.~Fraser, L.-P. Ho, J.~M. Wild, and F.~Gleeson, ``Hyperpolarized {129Xe} {MRI} {Abnormalities} in {Dyspneic} {Patients} 3 {Months} after {COVID}-19 {Pneumonia}: {Preliminary} {Results},'' {\em Radiology}, vol.~301, pp.~E353--E360, Oct. 2021.
\newblock Publisher: Radiological Society of North America.

\bibitem{deBoer:2010slp}
W.~de~Boer, J.~Lasenby, J.~Cameron, R.~Wareham, S.~Ahmad, C.~Roach, W.~Hills, and R.~Iles, ``Slp: A zero-contact non-invasive method for pulmonary function testing.,'' in {\em BMVC}, pp.~1--12, 2010.

\bibitem{ElshafieEtAl:2016sl}
G.~Elshafie, P.~Kumar, S.~Motamedi-Fakhr, R.~Iles, R.~C. Wilson, and B.~Naidu, ``{Measuring changes in chest wall motion after lung resection using structured light plethysmography: a feasibility study},'' {\em Interactive CardioVascular and Thoracic Surgery}, vol.~23, pp.~544--547, 06 2016.

\bibitem{MotamediFakhrEtAl:2017sl}
R.~C.~W. Shayan Motamedi-Fakhr and R.~Iles, ``Tidal breathing patterns derived from structured light plethysmography in copd patients compared with healthy subjects,'' {\em Medical Devices: Evidence and Research}, vol.~10, pp.~1--9, 2017.

\bibitem{JolliffeEtAl:2016rv}
I.~T. Jolliffe and J.~Cadima, ``Principal component analysis: a review and recent developments,'' {\em Philosophical Transactions of the Royal Society A: Mathematical, Physical and Engineering Sciences}, vol.~374, p.~20150202, April 2016.

\bibitem{Greenacre:2010rv}
M.~J. Greenacre, ``Correspondence analysis,'' {\em WIREs Computational Statistics}, vol.~2, pp.~613--619, 2024/08/27 2010.

\bibitem{Pages:2014fa}
J.~Pag{{\`e}}s, {\em Multiple Factor Analysis by Example Using R}.
\newblock Boca Raton, FL: Chapman and Hall/CRC, 2014.

\bibitem{LeEtAl:2008rp}
S.~L{\^e}, J.~Josse, and F.~Husson, ``Factominer: An r package for multivariate analysis,'' {\em Journal of Statistical Software}, vol.~25, pp.~1 -- 18, 2024/08/27 2008.

\bibitem{mcinnes_umap_2018}
L.~McInnes, J.~Healy, and J.~Melville, ``{UMAP}: {Uniform} {Manifold} {Approximation} and {Projection} for dimension reduction [{Preprint}],'' 2018.
\newblock Publisher: arXiv Version Number: 3.

\bibitem{alhuthail_measurement_2021}
E.~Alhuthail, J.~Stockley, A.~Coney, and B.~Cooper, ``Measurement of breathing in patients with post-{COVID}-19 using structured light plethysmography ({SLP}),'' {\em BMJ Open Respiratory Research}, vol.~8, p.~e001070, Oct. 2021.
\newblock Publisher: Archives of Disease in childhood Section: Respiratory physiology.

\bibitem{frey_temporal_2011}
U.~Frey, G.~Maksym, and B.~Suki, ``Temporal complexity in clinical manifestations of lung disease,'' {\em Journal of Applied Physiology}, vol.~110, pp.~1723--1731, June 2011.
\newblock Publisher: American Physiological Society.

\bibitem{aliverti_compartmental_2001}
A.~Aliverti, R.~Dellacà, P.~Pelosi, D.~Chiumello, L.~Gattinoni, and A.~Pedotti, ``Compartmental {Analysis} of {Breathing} in the {Supine} and {Prone} {Positions} by {Optoelectronic} {Plethysmography},'' {\em Annals of Biomedical Engineering}, vol.~29, pp.~60--70, Jan. 2001.

\bibitem{fleck2019}
D.~Fleck, C.~Curry, K.~Donnan, O.~Logue, K.~Graham, K.~Jackson, K.~Keown, J.~Winder, M.~D. Shields, and C.~M. Hughes, ``Investigating the clinical use of structured light plethysmography to assess lung function in children with neuromuscular disorders,'' {\em PLoS One}, vol.~14, no.~8, p.~e0221207, 2019.

\bibitem{Marconi_De_Lazzari_2020}
S.~Marconi and C.~De~Lazzari, ``In silico study of airway/lung mechanics in normal human breathing,'' {\em Mathematics and Computers in Simulation}, vol.~177, pp.~603--624, Nov. 2020.

\bibitem{jin2020}
J.-M. Jin, P.~Bai, W.~He, F.~Wu, X.-F. Liu, D.-M. Han, S.~Liu, and J.-K. Yang, ``Gender differences in patients with covid-19: focus on severity and mortality,'' {\em Frontiers in public health}, vol.~8, p.~545030, 2020.

\bibitem{MandelbrotWallis:1969lm}
B.~B. Mandelbrot and J.~R. Wallis, ``Robustness of the rescaled range r/s in the measurement of noncyclic long run statistical dependence,'' {\em Water Resources Research}, vol.~5, no.~5, pp.~967--988, 1969.

\bibitem{GrangerJoyeux:1980lm}
C.~W.~J. Granger and R.~Joyeux, ``An introduction to long-memory time series models and fractional differencing,'' {\em Journal of Time Series Analysis}, vol.~1, no.~1, pp.~15--29, 1980.

\bibitem{Hosking:1981fd}
J.~R.~M. Hosking, ``{Fractional differencing},'' {\em Biometrika}, vol.~68, pp.~165--176, 04 1981.

\bibitem{MarronDryden:2021oo}
J.~S. Marron and I.~L. Dryden, {\em Object oriented data analysis}.
\newblock Chapman and Hall/CRC, 2021.

\bibitem{Ward1992}
M.~E. Ward, J.~W. Ward, and P.~T. Macklem, ``Analysis of human chest wall motion using a two-compartment rib cage model.,'' {\em Journal of applied physiology (Bethesda, Md. : 1985)}, vol.~72, pp.~1338--47, Apr 1992.

\end{thebibliography}
\bibliographystyle{ieeetr}

\label{EndOfDoc}
\end{document}